\documentclass[seceq,preprint]{ptptex}

\usepackage{graphicx}
\usepackage{wrapft}
\usepackage{epsfig}
\def\vc#1{\mbox{\boldmath $#1$}}

\newcommand{\be}{\begin{equation}}
\newcommand{\ee}{\end{equation}}
\newcommand{\bea}{\begin{eqnarray}}
\newcommand{\eea}{\end{eqnarray}}

\begin{document}

\title{Alpha-clustered hypernuclei and chiral SU(3) dynamics}


\author{Emiko \textsc{Hiyama}$^a$%
,Yasuro \textsc{Funaki}$^a$, Norbert \textsc{Kaiser}$^b$, and Wolfram \textsc{Weise}$^{b,c}$}

\inst{
$^a$RIKEN Nishina Center, 2-1 Hirosawa, Wako 351-0106, Japan\\$^b$Physik-Department, Technische 
Universit\"at M\"unchen, D-85747 Garching, Germany\\
$^c$ECT$^*,$ Villa Tambosi, I-38123 Villazzano (Trento), Italy
}



\abst{
Selected light hypernuclei are studied 
using an $\alpha$ cluster model approach (the Hyper-THSR wave function) in combination with a
density-dependent $\Lambda$ hyperon-nuclear interaction derived from chiral $SU(3)$ 
effective field theory. This interaction includes important two-pion exchange
processes involving $\Sigma N$ intermediate states and associated three-body mechanisms 
as well as effective mass and surface terms arising in a derivative expansion of the 
in-medium $\Lambda$ self-energy. Applications
and calculated results are presented and discussed for $^{13}_\Lambda$C and $_\Lambda^9$Be.
The lightest $\alpha$ clustered hypernucleus, $^5_{\Lambda}$He, is also discussed in this context.
}

\maketitle

\section{Introduction}

The physics of $\Lambda$ hypernuclei has a long and well-established history. With the 
increasing precision of hypernuclear spectroscopy, accurate constraints on the 
effective interaction of the $\Lambda$ with nucleons in the nuclear medium emerge 
\cite{Tamura2010,Hiyama2009,Gal2008,Millener2008,Tamura2012}.  Empirical 
single-particle energies of a $\Lambda$ bound in hypernuclei  are well described in 
terms of an attractive mean field that is about half as strong ($U_\Lambda \approx 
-30\,$MeV) as the single-particle potential for nucleons in a nucleus. The empirical 
$\Lambda$-nuclear spin-orbit interaction is extremely  weak  compared to that of a  
nucleon in the nucleus. All these features have been described quite successfully, 
at least for medium-mass and heavy hypernuclei,  in various phenomenological 
mean-field approaches inspired by the shell-model picture.  In recent years an equally 
successful theoretical framework has been developed \cite{KW2005} based on chiral 
SU(3) dynamics, the effective field theory at the interface of three-flavor, 
low-energy QCD and nuclear physics with strangeness degrees of freedom. When 
converting the energy density derived in this framework into a hypernuclear density 
functional, this theory provides a quantitative description of $\Lambda$ hypernuclei 
over a wide mass range \cite{FKVW2009}, from $^{16}_\Lambda$O to $^{208}_\Lambda$Pb. 

For lighter hypernuclei such as $^{13}_\Lambda$C this approach works still reasonably well 
but turns out to be less accurate. A primary reason for this is the more complex 
structure of the corresponding core nuclei. Many previous investigations have shown that clustering correlations
play an important role in light nuclei \cite{WT1977,IHS1980}. A typical example is $^{12}$C which is known not to be 
a good shell-model nucleus. Its first excited $0^+$ state (the famous Hoyle state) has 
a pronounced cluster structure displaying a strong component of three alpha clusters 
in its wave function.  Such a structure also emerges in {\it ab initio} Monte Carlo lattice computations based
on chiral effective field theory \cite{epelbaum2012}. Recent works support the picture that 
the Hoyle state can, to a good approximation, be described as a product state of weakly interacting alpha particles occupying the lowest $0S$-orbit of a mean-field potential, with a relatively low average density, only about ${1\over 3}$ to 
${1\over 4}$ of the nuclear saturation density \cite{thsr,funaki_concept,funaki_12C}. The $^{12}{\rm C}$ ground state, while displaying a leading shell-model configuration, has nonetheless a pronounced component of three strongly 
correlated alpha clusters \cite{YFHIT2008}. Likewise, the $^8$Be nucleus features a prominent $\alpha\alpha$ clustering substructure. When a $\Lambda$ hyperon is added to the nuclear core,
significant changes of the ground state configuration can be induced as discussed e.g. in refs. 
\citen{Win2008,Isaka2011}.

The aim of the present study is to investigate the interaction of a $\Lambda$ 
with such clustered core nuclei, based on a density-dependent Hamiltonian derived from chiral 
SU(3) dynamics. Apart from a central potential, this interaction features a 
characteristic surface (derivative) coupling which is sensitive to the detailed shape of the nuclear 
density profile. This density distribution, and in particular its surface shape, is 
in turn influenced by the microscopic structure of the core wave function. The primary focus in this investigation is on $_\Lambda^{13}$C, while calculations are also performed for $_\Lambda^9$Be, 
starting from realistic dynamical cluster wave functions of their $^{12}$C and $^{8}$Be cores. 
Since the relative proportions of surface and volume change significantly in these two 
nuclei, they provide a testing ground for a detailed study of the interplay 
between bulk and surface terms in the chiral SU(3) based $\Lambda$-nuclear interaction. 

For the lightest $\Lambda\alpha$ compound, the $^5_{\Lambda}$He hypernucleus, the question
arises whether the present effective $\Lambda$-nuclear interaction and its 
pronounced derivative term apply also to this more compact system. Calculations of
the $^5_{\Lambda}$He binding energy have so far frequently used a schematic Gaussian type
density for the $\alpha$ core, with its size parameter fixed to reproduce the observed 
charge radius of $^4$He. However, in a recent four-body calculation \cite{HGK2004}
it was found that the strong $NN$ correlations in $^4$He imply a significant deviation of the 
resulting density from a Gaussian form, with consequences also for the detailed density profile and
its surface. It is therefore of interest to calculate, in addition, the binding energy of $^5_{\Lambda}$He 
using the wave function of $^4$He generated by the four-body calculation.

This paper is organized as follows. Section 2 briefly summarizes the $\Lambda$-nuclear interaction derived from chiral SU(3) dynamics and employed in this work. In Sections 3 and 4, the model wave functions and density distributions for $^{13}_{\Lambda}$C and $^9_{\Lambda}$Be are introduced. Results and discussions are presented in Section 5 followed by a summary in Section 6.

\section{The $\Lambda$-nuclear interaction from chiral SU(3) dynamics}

In previous work \cite{KW2005,FKVW2009}, the interaction of a $\Lambda$ hyperon with a 
nuclear medium has been derived using the chiral SU(3) meson-baryon effective 
Lagrangian at next-to-leading order (NLO) as a starting point. An important element 
of this approach is the systematic treatment of kaon and two-pion exchange processes 
governing the in-medium $\Lambda$N interactions. While direct single-pion exchange in 
the $\Lambda$N system is isospin-forbidden, iterated pion exchange driven by the 
second-order tensor force and involving an intermediate $\Sigma$ hyperon provides 
the dominant mid-range attraction. Short-distance dynamics, not resolved at the 
relevant nuclear Fermi momentum scales, are encoded in a
few contact terms with coefficients adjusted to reproduce bulk properties of 
hypernuclei. The remaining parameters of the theory are the known structure constants 
of the pseudoscalar meson octet (the pion and kaon decay constants in vacuum) and the 
axial vector coupling constants of the baryon octet (determined by nucleon and hyperon 
beta decays). A calculation of all NLO contributions has been performed at two-loop 
order for the $\Lambda$-nuclear (central and spin-orbit) mean fields, with full 
account of important Pauli-blocking effects in the nuclear medium \cite{KW2005}. 

Within this framework, the self-energy of the $\Lambda$ interacting with the nuclear 
many-body system has been constructed.  Its dependence on the nuclear density, $\rho_N = 
2k_F^3/3\pi^2$, can be represented in the form of an expansion in powers of the 
nucleon Fermi momentum, $k_F$. In a subsequent step this self-energy has been 
translated into a $\Lambda$-nucleus potential $U_\Lambda({\bf r})$ for applications 
to hypernuclei, using a derivative expansion in terms of the local density, 
$\rho_N(\bf r)$, of the nuclear core. The result is as follows \cite{KW2005}:
\begin{equation}
U_\Lambda({\bf r}) = U_c({\bf r}) - {1\over 2M_\Lambda}\vec{\nabla}\cdot R(\rho_N)
\vec{\nabla} - D(\rho_N)\big(\vec{\nabla}^2\rho_N({\bf r})\big)~~,
\label{eq:lambdapot}
\end{equation}
where the central part is written as an expansion in fractional powers of $\rho_N$:
\begin{eqnarray}
U_c({\bf r}) &=& 
U_0\,{\rho_N({\bf r})\over \rho_0}\,\times \nonumber \\
&& \left[1 + 0.351\left(\rho_N
({\bf r})\over \rho_0\right)^{1/3}-0.359\left(\rho_N({\bf r})\over \rho_0\right)^{2/3}
-0.033\left(\rho_N({\bf r})\over \rho_0\right)\right]~~,
\label{eq:lambdacentral}
\end{eqnarray}
with $U_0 = -30.56$ MeV, about half the strength of the single-particle potential for 
nucleons in nuclei. A slightly more attractive potential, $U_0 \simeq -35$ MeV, was found in the 
systematic analysis of hypernuclear binding energies using a similar approach \cite{FKVW2009}.  
In the expansion Eq. (\ref{eq:lambdacentral}) the core density $\rho_N$ is expressed in units of 
normal nuclear matter density, $\rho_0 = 0.16$ fm$^{-3}$. 
The leading term linear in $\rho_N$ 
is characteristic of the Hartree mean field approximation. Non-trivial terms beyond 
this linear density dependence arise from two-pion exchange dynamics 
in the medium with inclusion of Pauli-blocking effects.  

The derivative terms in Eq.(\ref{eq:lambdapot}) reflect the momentum 
dependence of the $\Lambda$ self-energy in the nuclear medium at order $p^2$. 
In r-space these derivative terms represent non-local effects beyond the simplest local 
density approximation. Such terms are expected to become increasingly important
as the proportion of surface to bulk increases in light nuclei. The piece proportional to 
$R(\rho_N)$ contributes to the (density dependent) effective mass of the 
$\Lambda$ hyperon. When combined with the kinetic energy piece of the (free) 
$\Lambda$ Hamiltonian, one has
\begin{equation}
H_{\Lambda,kin} = - {1\over 2M_\Lambda}\vec{\nabla}\cdot \left[1 + R(\rho_N)\right]
\vec{\nabla} ~~.\label{eq:lambdakin} \end{equation}
The explicit analytical expression for $R(\rho_N)$ can be found in the appendix. For 
the present purpose it is well approximated by the series:
\begin{eqnarray}
R\left(\rho_N({\bf r})\right) &=& -0.073\left(\rho_N({\bf r})\over \rho_0\right) 
-0.098\left(\rho_N({\bf r})\over \rho_0\right)^{4/3} \nonumber \\
&& -0.101\left(\rho_N({\bf r})\over 
\rho_0\right)^{5/3}+0.056\left(\rho_N({\bf r})\over \rho_0\right)^2~~.\label{eq:R}
\end{eqnarray}
At typical densities, $\rho_N \sim \rho_0/2$, this gives a small but significant 
correction to the effective $\Lambda$ mass, $M_\Lambda^*(\rho_N)/M_\Lambda = \left[1+ 
R(\rho_N)\right]^{-1}$, of about 10\%. 

The third term in Eq.(\ref{eq:lambdapot}), the one
proportional to the Laplacian acting on the density, is sensitive to the detailed 
surface profile of $\rho_N({\bf r})$. The analytical expression for $D(\rho_N)$ can again be 
found in the appendix. From previous analyses of intermediate-mass and 
heavy hypernuclei \cite{FKVW2009}, the surface coupling strength $D(\rho_N)$ turns 
out to be approximately constant, i.e. independent of density. Values of $D$ are in the 
range $D \sim (0.2 - 0.4)$ fm$^4 \simeq (40 - 80)$ MeV$\cdot$ fm$^5$ or smaller 
depending on the attractive strength of the central (local) $\Lambda$-nuclear mean field.
While it is not possible to determine the surface coupling strength $D$ more accurately from 
heavier hypernuclei for which the fraction of surface-to-volume is small, it is 
of interest to analyse in more detail the interplay between the strengths $U_0$ and $D$ of the 
central and surface potentials, respectively, for lighter hypernuclei. This is the primary task 
of the present study which combines the input $\Lambda$-nuclear interaction with a
nuclear core wave function constructed from a microscopic cluster model.

\section{Hypernuclear alpha cluster structure}

Consider as a starting point a Hamiltonian for $N=Z$ nuclei with $4n$ nucleons ($n=2$ 
for Be and $n=3$ for C) plus a $\Lambda$ hyperon, composed of kinetic energies $-\frac{1}{2M}\vec \nabla_i^2$ 
(with nucleon mass $M$) and $-\frac{1}{2M_\Lambda}\vec \nabla_\Lambda^2$ (with $\Lambda$ 
hyperon mass $M_\Lambda$), the Coulomb potential $V_{ij}^C$, the effective 
nucleon-nucleon interaction $V^{NN}_{ij}$, and the $\Lambda$-nucleon $(\Lambda N)$ 
interaction $V^{\Lambda N}_{i}$:
\begin{equation}
H=-\sum_{i=1}^{4n}\frac{1}{2M}\vec \nabla_i^2 -\frac{1}{2M_\Lambda}\vec \nabla_\Lambda^2 
- T_G +\sum_{i<j}^{4n}V_{ij}^C + \sum_{i<j}^{4n} V^{NN}_{ij} + \sum_{i=1}^{4n}V_{i}^{\Lambda N}~. 
\label{eq:hml} \end{equation}
The center-of-mass kinetic energy $T_G$ is properly subtracted. We neglect the small 
$\Lambda N$ spin-orbit interaction. In the actual calculation the Volkov No.2 
NN-force~\cite{volkov} for ${^8{\rm Be}}$ is used, and a slightly modified version of this 
force \cite{kamimura} for ${^{12}{\rm C}}$. The $\Lambda N$ interaction is provided 
by the phenomenological Nijmegen potential (model D) 
\cite{HKM1997} for the purpose of computing  wave functions and density 
distributions for ${^9_\Lambda{\rm Be}}$ and ${^{13}_\Lambda{\rm C}}$. With this phenomenological input the calculation of the $\Lambda$ binding energy yields 6.69\,MeV (exp.: 6.71\,MeV \cite{expBe}) for 
${^9_\Lambda{\rm Be}}$, and 11.68\,MeV (exp.: 11.71\,MeV \cite{expC}) for 
${^{13}_\Lambda{\rm C}}$, respectively. Ultimately these microscopic cluster calculations should 
be updated replacing phenomenological forces by new baryon-baryon interactions derived consistently 
from chiral dynamics \cite{Haidenbauer2013}.

In the present work the focus is on the interaction of the $\Lambda$ with 
the nuclear core based on chiral SU(3) effective field theory, replacing
\begin{equation}
H_\Lambda = -\frac{1}{2M_\Lambda}\vec \nabla^2_\Lambda + \sum_{i=1}^{4n}V_{i}^{\Lambda N} ~~ 
\rightarrow~~  - {1\over 2M_\Lambda}\vec{\nabla}\cdot \left[1 + R(\rho_N)\right]
\vec{\nabla} + U_c({\bf r}) - D\,\big(\vec{\nabla}^2\rho_N({\bf r})\big)~~, 
\label{eq:hmlambda}
\end{equation}
where the gradients in the first term are understood to act on the $\Lambda$ hyperon 
coordinate and the remaining expressions are as specified in the previous section. 
Taking expectation values of this new interaction with calculated wave functions, 
the aim is then to study in particular the role of the genuine surface term of 
Eq.(\ref{eq:hmlambda}). The effect of the derivative term in light hypernuclei is examined 
here for the first time. The importance of this term has been established in previous calculations 
for a $\Lambda$ in slightly inhomogeneous nuclear matter \cite{KW2005} and for hypernuclei 
ranging from $^{16}_\Lambda{\rm O}$ to $^{208}_\Lambda{\rm Pb}$ \cite{FKVW2009}.

\section{Derivation of the nuclear core density}
The quantity of key importance is now the nuclear core density distribution 
$\rho_N({\bf r})$ in the hypernucleus. The model wave function used here to calculate 
this density is the so-called Hyper-THSR (Tohsaki-Horiuchi-Schuck-R\"opke) wave function.
It is based on the deformed 
THSR wave function~\cite{thsr,funaki_8Be} describing nuclei with $4n$ nucleons as follows:
\begin{equation}
 \Phi_{n\alpha}^{\rm THSR}(\vc{B}) \propto {\cal A}\ \Big\{ \prod_{i=1}^n\exp \Big[- 
\sum_{k=x,y,z}\frac{2}{B_k^2} (X_{ik}-X_{Gk})^2 \Big] \phi(\alpha_i) \Big\}, 
 \label{eq:thsr} 
\end{equation}
with the antisymmetrizer $\cal A$ operating on all nucleons and $\phi(\alpha_i)$ the 
intrinsic wave function of the $i$-th $\alpha$ cluster:
\begin{equation}
\phi(\alpha_i) \propto \exp\Big[-\sum_{1\leq k<l \leq4}({\vc r}_{i,k} - 
{\vc r}_{i,l})^2/(8b^2)\Big]~~.\label{eq:int_alpha}
\end{equation}
In Eq.\,(\ref{eq:thsr}), $\vc{X}_i$  denotes the center-of-mass coordinates of the 
$i$-th $\alpha$ particle. The spurious total center-of-mass coordinate $\vc{X}_G$ is 
properly eliminated. The center-of-mass motions of the $n$ alpha clusters occupy the 
same deformed orbit, $\exp [-\sum_{k=x,y,z}$ $\frac{2}{B_k^2}(X_k-X_{Gk})^2 ]$, 
displaying a product arrangement of the $n\, \alpha$ particles when $\vc{B}$ is so large that the 
effect of the antisymmetrizer becomes negligible~\cite{funaki_concept}. In the 
limiting case $B_k \rightarrow \infty \ (k=x,y,z)$, this wave function corresponds to 
a free $n\alpha$ state in which the $\alpha$ particles are uncorrelated. In the 
symmetric limit $B_x=B_y=B_z \rightarrow b$ the normalized THSR wave function 
coincides with the shell model Slater determinant. 

The wave function, Eqs.(\ref{eq:thsr},\ref{eq:int_alpha}), succeeded in describing 
light $N=Z$ nuclei such as ${^{8}{\rm Be}}$, ${^{12}{\rm C}}$, 
${^{16}{\rm O}}$ and ${^{20}{\rm Ne}}$. In particular the first excited $0^+$ state in 
${^{12}{\rm C}}$ is described correctly by this wave function, with a loosely coupled 
structure of the three $\alpha$ particles reminiscent of a gas occupying the 
lowest $0S$ orbit of a mean-field potential for the $\alpha$ 
particle~\cite{funaki_12C}. Apart from this example, the THSR wave function is known 
to give a valid description of the ground states of those nuclei since their shell 
model configurations are properly taken into account due to the antisymmetrization of 
nucleons, together with the $\alpha$-like ground state correlations. 
In particular, for the ground state rotational bands of $^{12}{\rm C}$ and 
$^{20}{\rm Ne}$, the THSR wave functions~\cite{funaki_concept, zhou_bo} give large squared overlap 
(close to 100 \%) with the corresponding microscopic cluster model wave functions 
such as RGM (Resonating Group Method) and GCM (Generator Coordinate 
Method)~\cite{carbon}.

The Hyper-THSR wave function describing the $4n+\Lambda$ hypernuclei is then 
introduced as follows:
\begin{equation}
 \Phi_{n\alpha-\Lambda}^{\rm H-THSR} (\vc{B},\kappa) =  \Phi_{n\alpha}^{\rm THSR}(\vc{B}) 
\varphi_\Lambda(\kappa), \label{eq:hypthsr}
\end{equation}
where the $\Lambda$ particle is assumed to couple to the $n\alpha$ core nucleus in an $S$ 
wave. This approximation is supported by previous calculations \cite{hiyama2000} for $^{13}_\Lambda{\rm C}$ and $^{9}_\Lambda{\rm Be}$ where it was found that the ground states of these hypernuclei are dominated (to 94.4 \% for $^{9}_\Lambda{\rm Be}$ and 98.8 \% for  $^{13}_\Lambda{\rm C}$) by configurations with the $\Lambda$ in an s-orbit coupled to a nuclear $0^+$ core. This also underlines the justification for an approximate treatment of such systems as effective two-body problems. 

The radial part of the wave function Eq. (\ref{eq:hypthsr}) is 
expanded in Gaussian basis functions, $\varphi_\Lambda(\kappa)=(\pi/2\kappa)^{-3/4}
\exp(-\kappa r_{n\alpha-\Lambda}^2)$. In the practical calculations we use an 
axially-symmetric function in Eq.~(\ref{eq:thsr}) with $B_x=B_y\equiv B_\perp \ne B_z$.
In this way the intrinsic deformation of the core wave function is taken into account. 
The parameter $b$ in Eq.~(\ref{eq:int_alpha}) is fixed to almost the same size as the one of  
the $\alpha$ particle in free space. The total wave function for quantum 
states of the $n\alpha + \Lambda$ nucleus can then be expressed as the superposition of the 
angular-momentum projected wave function of Eq.~(\ref{eq:hypthsr}) with different 
values of the parameters, $B_\perp, B_z$ and $\kappa$, as follows:
\begin{equation}
\Psi_{n\alpha-\Lambda}^{\rm H-THSR} (J_\lambda^+) = \sum_{B_\perp,B_z,\kappa}f_\lambda (B_\perp,B_z,
\kappa) {\hat P}^{J} \Phi_{n\alpha-\Lambda}^{\rm H-THSR} (B_\perp, B_z,\kappa), \label{eq:qwf}
\end{equation}
where ${\hat P}^J$ is the angular-momentum projection operator onto the $J^\pi$subspace
with positive parity, $\pi=+$ (note that the Hyper-THSR wave function, Eq.(\ref{eq:hypthsr}), has positive intrinsic 
parity). The coefficients $f_\lambda(B_\perp,B_z,\kappa)$ are then determined by solving 
the following Griffin-Hill-Wheeler equation~\cite{GHW}: 
\begin{equation}
\sum_{B_\perp^\prime,B_z^\prime,\kappa^\prime} \big\langle \Phi_{n\alpha-\Lambda}^{\rm H-THSR} 
(B_\perp,B_z,\kappa) \big|{\hat H}-E^\lambda \big|{\hat P}^J \Phi_{n\alpha-\Lambda}^{\rm H-THSR} 
(B_\perp^\prime,B_z^\prime,\kappa^\prime) \big\rangle f_\lambda (B_\perp^\prime,B_z\prime,
\kappa^\prime)=0. \label{eq:ghweq}
\end{equation}
Using the wave function (\ref{eq:qwf}) the (radial) nucleon density distribution of 
the core, averaged over angular dependence, is introduced as: 
\begin{equation}
\rho_N(r)=\big\langle \Psi_{n\alpha-\Lambda}^{\rm H-THSR} (J_\lambda^+) \big| \frac{1}
{4\pi r^2}\sum_{i=1}^{4n}\delta(r - |\vc{r}_i-\vc{X}_C|) \big| 
\Psi_{n\alpha-\Lambda}^{\rm H-THSR} (J_\lambda^+) \big\rangle, \label{eq:dsty}
\end{equation}
where $\vc{X}_C=(\vc{r}_1+\cdots +\vc{r}_{4n})/(4n)$ is the center-of-mass 
coordinate of the $n\alpha$ core nucleus. Note again that the wave functions entering Eq.(\ref{eq:dsty})
are angular-momentum projected.
The nuclear core density $\rho_N$ is normalized as usual to the total number of nucleons, 
$\int d^3r\,\rho_N({\bf r}) = 4n$. 

\begin{figure}[htbp]
\begin{center}
\includegraphics[scale=0.3]{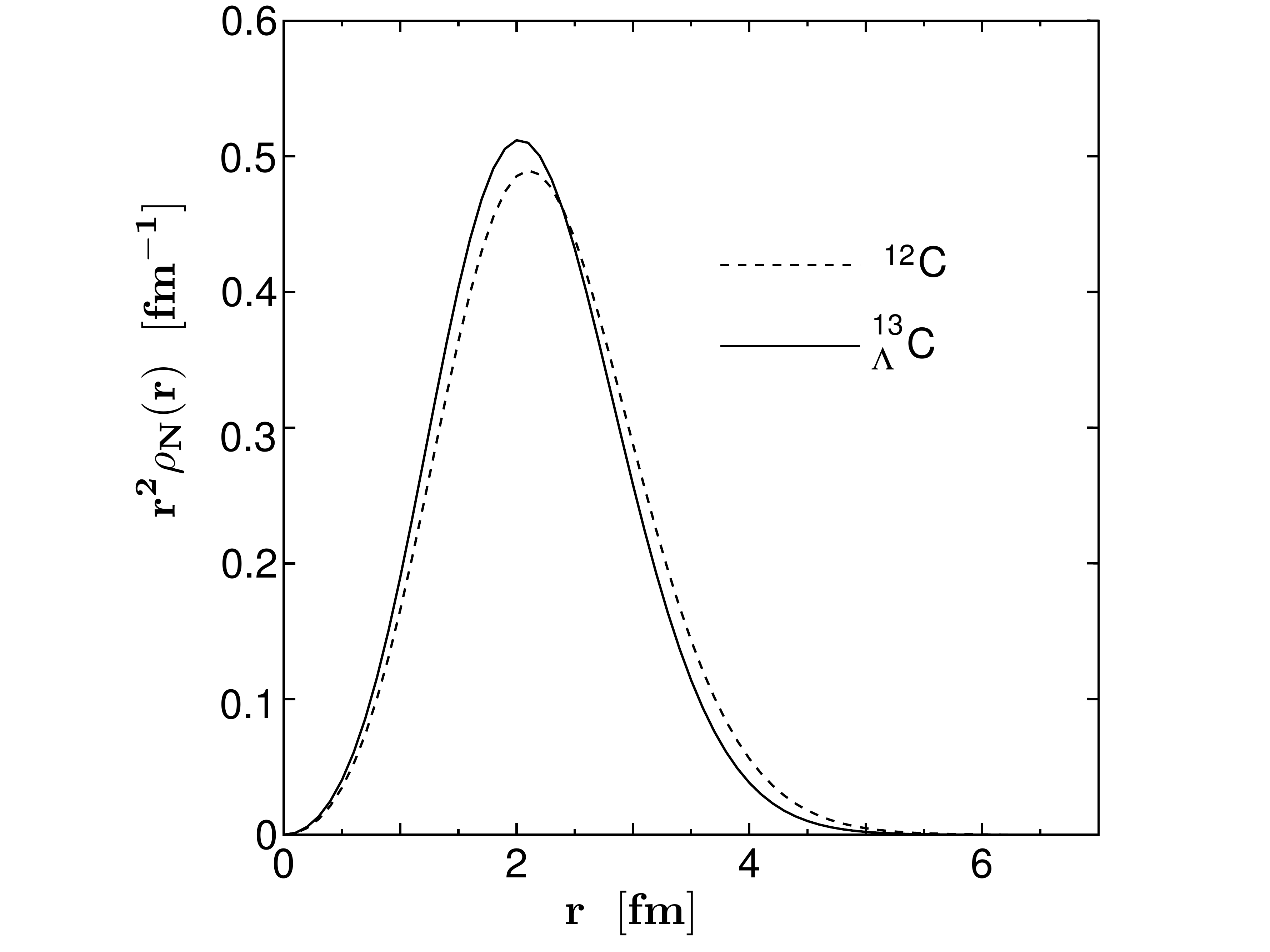}
\caption{Nucleon density distribution (multiplied by $r^2$) of ${^{13}_\Lambda{\rm C}}$ defined by 
Eq.~(\ref{eq:dsty}) (solid curve). For comparison, the density of the ${^{12}{\rm C}}$ 
nucleus is also shown by the dotted curve.}
\label{fig:dsty_13LC}
\end{center}
\end{figure}

\begin{figure}[htbp]
\begin{center}
\includegraphics[scale=0.3]{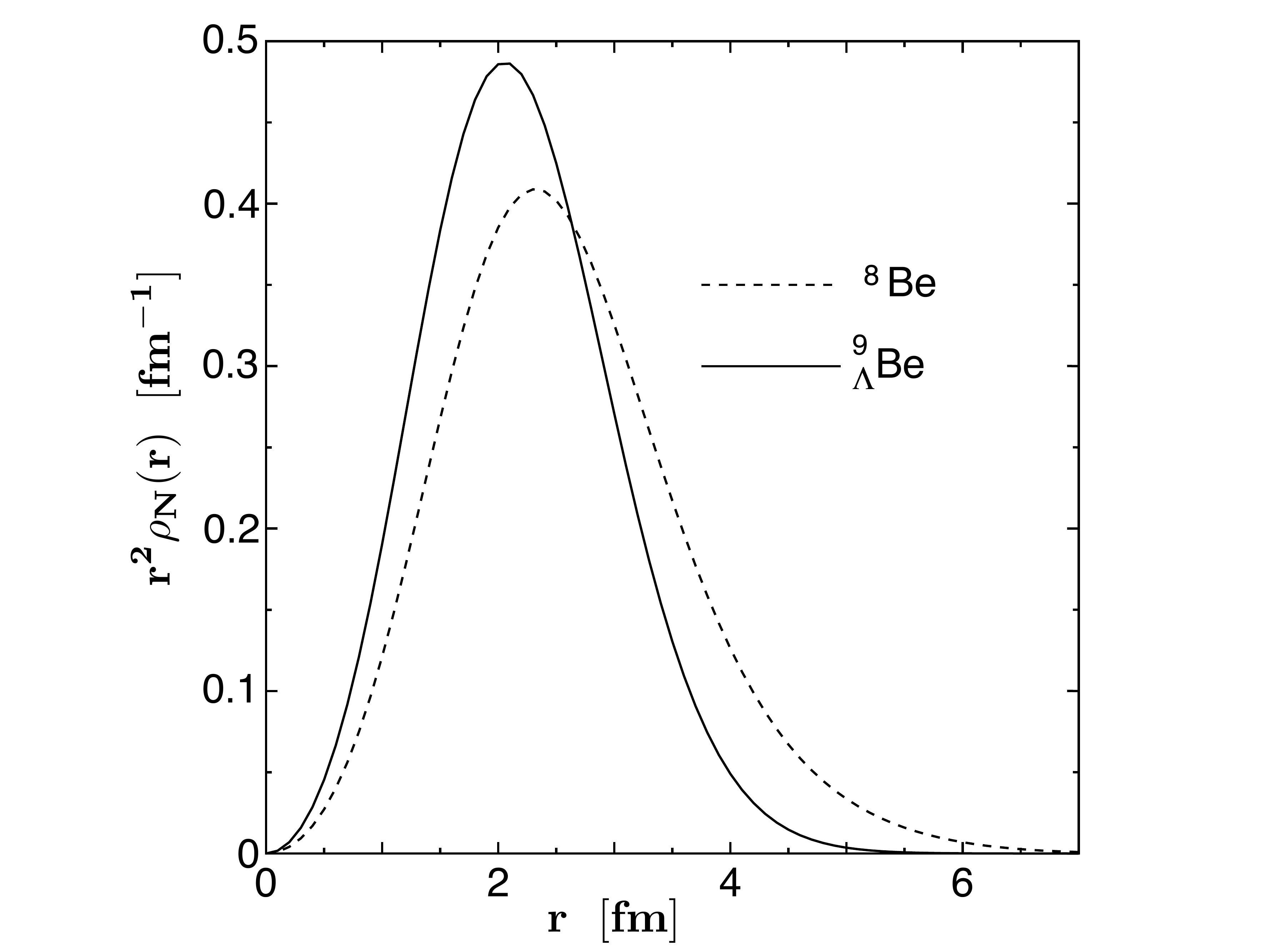}
\caption{Nucleon density distribution (multiplied by $r^2$) of ${^9_\Lambda{\rm Be}}$ defined by 
Eq.~(\ref{eq:dsty}) (solid curve). For comparison, the density of ${^8{\rm Be}}$ is also 
shown by the dotted curve.}
\label{fig:dsty_9LBe}
\end{center}
\end{figure}

In Figs.~\ref{fig:dsty_13LC} and \ref{fig:dsty_9LBe} we show $r^2$ times the density 
distributions $\rho_N$ of the nuclear core for the ground states of 
${^{13}_\Lambda{\rm C}}$ and ${^9_\Lambda{\rm Be}}$, i.e. for $n=3, J=0$ and $n=2, J=0$ 
in Eq.~(\ref{eq:dsty}), respectively. In both figures, the nucleon density 
distributions of the ground states of ${^8{\rm Be}}$ and ${^{12}{\rm C}}$ are  
shown for comparison. They are also obtained by solving the Griffin-Hill-Wheeler 
equation based on the $3\alpha$ or $2\alpha$ THSR wave functions. The calculated $^{12}{\rm C}$
charge density actually compares quite well with the empirical density deduced from electron scattering
data \cite{devries1987}.  
For ${^9_\Lambda{\rm Be}}$ one observes that a significant spatial shrinkage is induced by injecting the 
$\Lambda$ particle into ${^8{\rm Be}}$. The calculated root mean square (rms) radius for the 
${^8{\rm Be}}$ core in ${^9_\Lambda{\rm Be}}$ is $2.35$ fm, almost 20\% smaller than 
that for the ${^8{\rm Be}}$ nucleus ($2.87$ fm) which has a pronounced 
 $2\alpha$ cluster structure. This shrinkage effect is far less significant for 
${^{13}_\Lambda{\rm C}}$: the density of the ${^{12}{\rm C}}$ core in the hypernucleus 
is close to the normal density of the ${^{12}{\rm C}}$ nucleus. The calculated rms 
radius for the ${^{12}{\rm C}}$ core in ${^{13}_\Lambda{\rm C}}$ is $2.32$ fm, not much 
different from the rms radius ($2.40$ fm) of the ${^{12}{\rm C}}$ nucleus, but this difference is nonetheless
of some significance concerning the effect of the $\nabla^2\rho_N$ term.

\section{Results and Discussion}

Given the density distributions introduced in the previous section, we can now focus on the 
detailed investigation of the $\Lambda$-nuclear derivative coupling terms. Our primary example 
is $^{13}_{\Lambda}$C for which a direct comparison with the mean-field calculation of 
Ref.\citen{FKVW2009} is at hand and the derivative expansion with a local density $\rho_N$ is 
expected to work. For even lighter nuclei one expects larger uncertainties as the limits of 
applicability of this expansion may be encountered.

Consider the expectation values of the $\Lambda$ Hamiltonian in Eq.(\ref{eq:hmlambda}) 
taken with the correlated THSR wave functions for $^{13}_{\Lambda}$C and $^9_{\Lambda}$Be, 
using the $\Lambda$-nuclear interaction in Eq.(\ref{eq:lambdapot}) derived from chiral 
SU(3) effective field theory. We employ the nuclear core densities, $\rho_N({\bf r})$, 
determined relative to the center of mass of the $^{12}$C and $^8$Be cores in
$^{13}_{\Lambda}$C and $^9_{\Lambda}$Be, respectively. As an exploratory sideline, 
the $^5_{\Lambda}$He prototype hypernucleus will also be discussed.

We first study the case of  $^{13}_{\Lambda}$C. Its ground state is considered to be a compact
shell-model-like configuration for which the previously introduced $\Lambda$-core interaction 
is supposed to work well. It has been pointed out \cite{Yamada85,HKM1997} that
the density distribution of a compact state such as $^{13}_{\Lambda}$C 
does not experience a dynamical contraction when adding the $\Lambda$ particle to the nuclear
core. This feature is in fact realized in the present calculation, as seen in Fig. \ref{fig:dsty_13LC}.

Let us now examine the $\Lambda$ interaction with the $^{12}$C core in more detail. For example, with only 
the central piece $U_c$ of the interaction and choosing $U_0 = -35$\,MeV, the resulting calculated $\Lambda$ 
binding energy in $^{13}_{\Lambda}$C is 15.3 MeV. In the next step, including the second term of 
Eq.(\ref{eq:lambdapot}) that contributes to the effective mass $M_\Lambda^*$, the 
kinetic energy is reduced by slightly less than 10\% and so the binding increases to 
$B_\Lambda = 16.7$\,MeV. This overbinding is then compensated by the (repulsive) surface 
term proportional to $\nabla^2 \rho_N$. Choosing a surface coupling constant 
$D = 54$\,MeV$\cdot$fm$^5$ brings $B_\Lambda$ back to its empirical value, 
11.7\,MeV.

\begin{figure}[htbp]
\begin{center}
\includegraphics[scale=0.25]{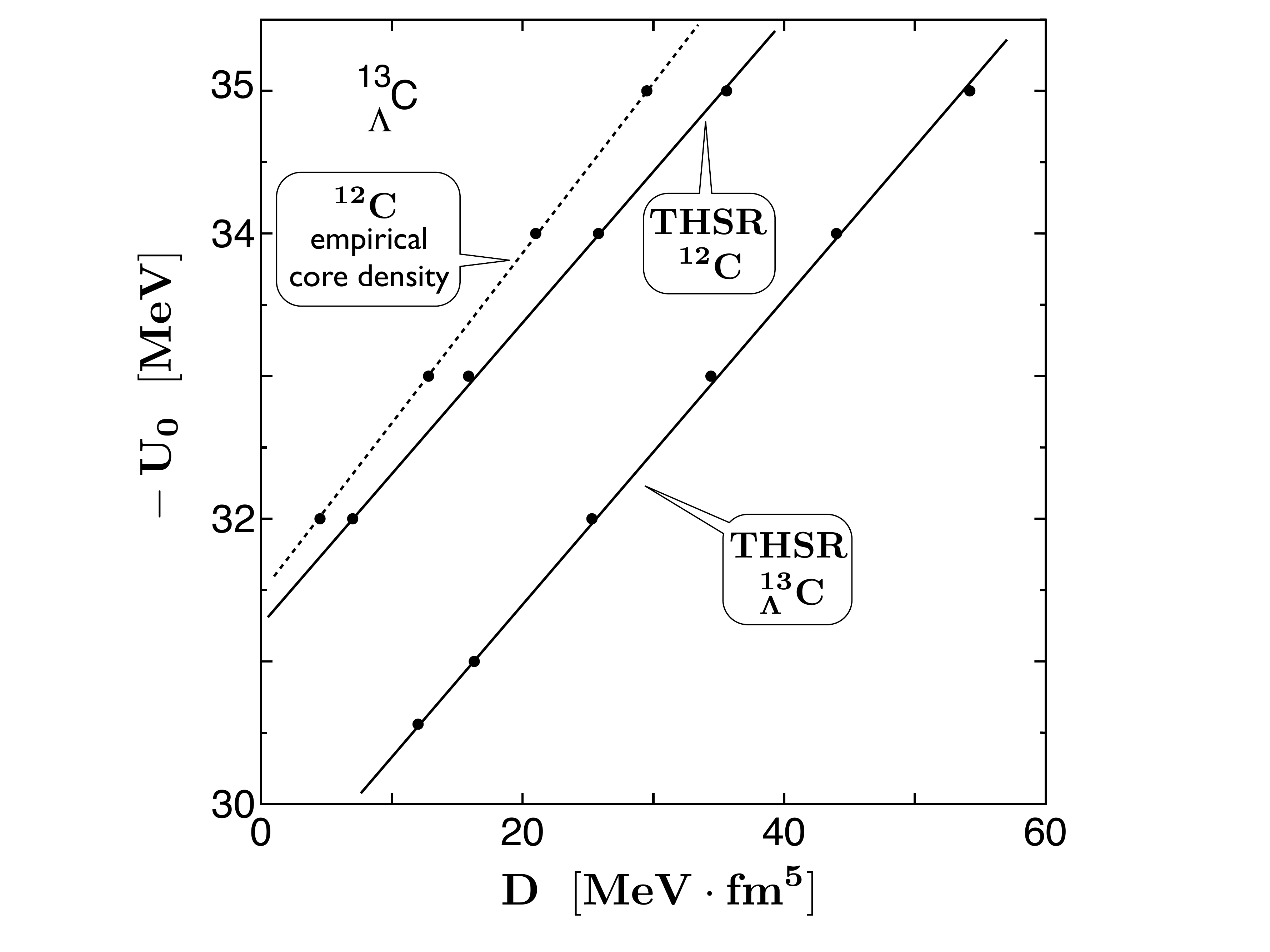}
\caption{Correlation between the $\Lambda$-nuclear potential depth $U_0$ and the 
strength $D$ of the surface term for $^{13}_\Lambda$C. The straight lines connect points that reproduce 
the empirical $\Lambda$ binding energy $B_\Lambda = 11.71$ MeV. The following cases using different A = 12 nuclear core densities $\rho_N$ are displayed in comparison: full hypernucleus calculation using Eq.~(\ref{eq:dsty}) (THSR $^{13}_\Lambda$C); calculated $^{12}$C core density using THSR cluster wave function (THSR $^{12}$C); empirical $^{12}$C core density deduced from electron scattering data (dashed curve). }
\label{fig:dfac}
\end{center}
\end{figure}

Fig.\ref{fig:dfac} demonstrates a subtle balance between the depth of the central potential, $U_0$, and 
the surface coupling strength $D$. Pairs of values $(U_0, D)$ that reproduce the empirical 
$B_\Lambda = 11.71$\,MeV are correlated linearly as shown in this figure. For example, the combination 
of $U_0=-32$ MeV with a surface term $D=25$ MeV$\cdot$fm$^5$ fits the empirical $^{13}_{\Lambda}$C binding 
energy of the $\Lambda$ equally well. From Ref. \citen{FKVW2009} we recall that, in the chiral $SU(3)$ effective field theory approach, the surface coupling strength $D(\rho_N)$ deduced from best-fit mean-field results for a variety of heavier hypernuclei ranges between about 40 and 80 MeV fm$^5$. The present analysis is consistent with such values if the strength of the central piece $U_c$ lies in the range $U_0 = - 33$ to $- 35$\,MeV. This result is based on the nuclear core density distribution Eq.  (\ref{eq:dsty}) as it emerges from the full calculation of the hypernuclear THSR wave function. For comparison, Fig.\ref{fig:dfac} also shows a study in which the nuclear A = 12 core density in $^{13}_{\Lambda}$C hypernucleus is simply replaced by the density distribution $\rho_N$ of the free, isolated $^{12}$C nucleus, either calculated using the THSR cluster model or using a parameterization of the empirical $^{12}C$ charge density \cite{devries1987}. In this case the suggested values of the surface coupling $D$ would be reduced. One observes that the relatively small difference seen in Fig.\ref{fig:dsty_13LC} between the density distribution of $^{12}$C and the nuclear core in $^{13}_{\Lambda}$C has nonetheless a pronounced effect on the derivative term proportional to $\nabla^2\rho_N$. 

\begin{figure}[htbp]
\begin{center}
\includegraphics[scale=0.25]{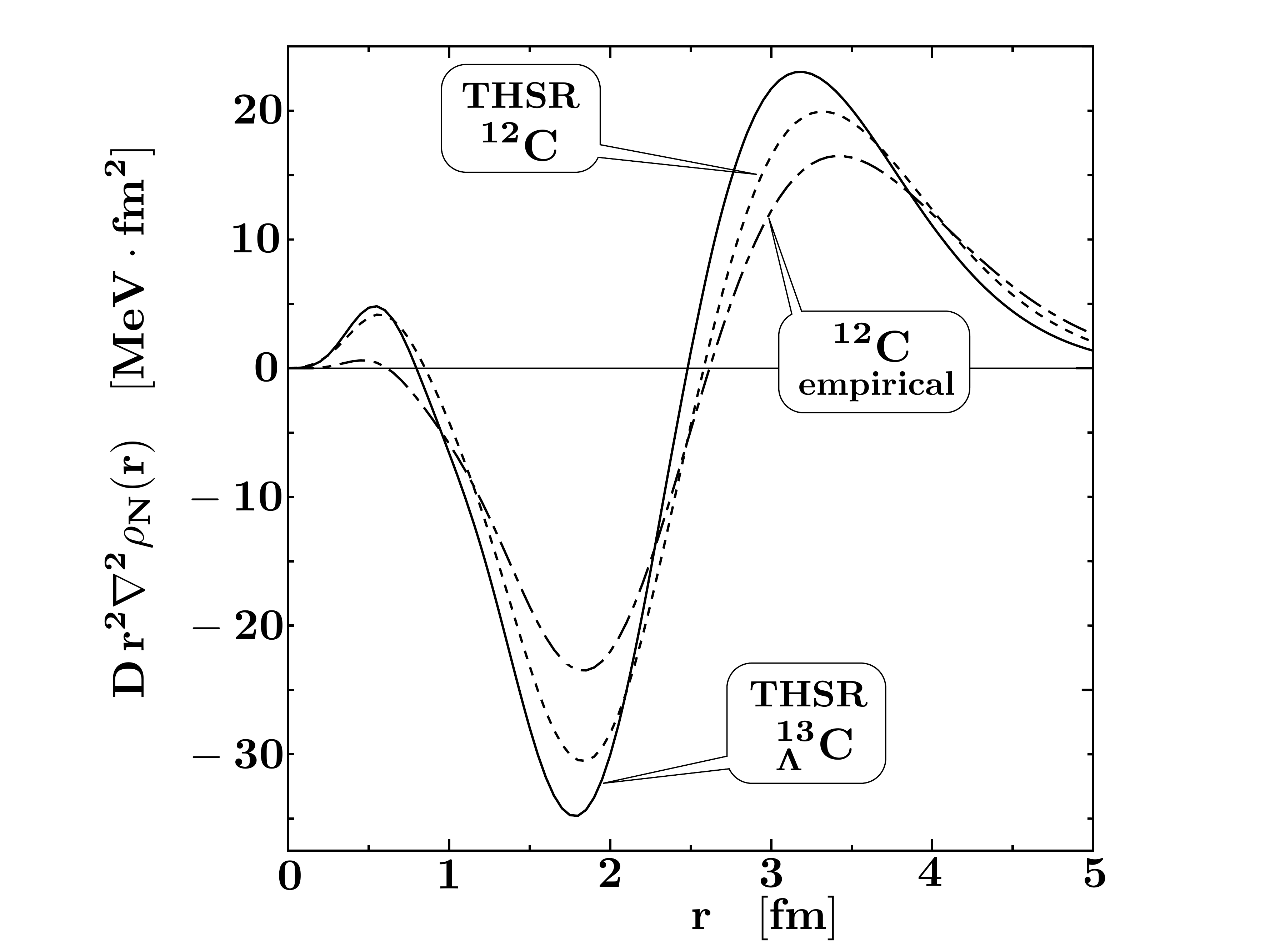}
\caption{Derivative term $D\nabla^2\rho_N(r)$ multiplied by $r^2$ (with $D = 50$ MeV$\cdot$fm$^5$). 
Solid curve: full  calculation of $^{13}_\Lambda$C using Eq.~(\ref{eq:dsty}) (THSR $^{13}_\Lambda$C); dashed curve: calculated $^{12}$C core density using THSR cluster wave function (THSR $^{12}$C); dash-dot curve: using empirical $^{12}$C core density deduced from electron scattering data. }
\label{fig:derivterm}
\end{center}
\end{figure}

The r-dependence of the $D\nabla^2\rho_N(r)$ term resulting from the $^{13}_{\Lambda}$C core density is displayed in Fig.\ref{fig:derivterm} together with the one from the calculated $^{12}$C density and in comparison with the one derived from the empirical $^{12}$C density distribution. The sensitivity of this derivative term (multiplied by $r^2$ as it appears in the relevant integral) with respect to the detailed behaviour of $\rho_N(r)$ is evident. Note that when combined with the square of the $\Lambda$ wave function in $^{13}_{\Lambda}$C, the weight in the relevant matrix element is dominantly in the range $r \sim 1-3$ fm, resulting in a net repulsive correction to the $\Lambda$ binding energy. It is also instructive to examine in more detail the effect of this derivative term on the $\Lambda$ binding energy with varying strength parameter $D$ as displayed in Table \ref{table:dfac}. 

\begin{table}[htb]
\caption{Binding energies (in MeV) of  $^{13}_{\Lambda}C$ 
with varying derivative coupling strength $D$ (in MeV$\cdot$fm$^5$) 
and central potential $U_0$ ranging between -32 and -35 MeV.}

\label{table:dfac}
\begin{center}
\begin{tabular}{rrrrrrrr}
\hline 
\vspace{-3 mm} \\
$^{13}_{\Lambda}$C &$~U_0$&~~~~  $D=0$  &$D=30$  &$D=50$ \\
\hline 
\vspace{-3 mm} \\
&-32  &14.15   &11.29   &9.47  \\
&-33  &15.00   &12.10   &10.44  \\
&-34  &15.85   &12.92   &11.23  \\
&-35  &16.70   &13.75   &12.04 \\
\hline \\

\end{tabular}
\end{center}
\end{table}

\begin{figure}[htbp]
\begin{center}
\includegraphics[scale=0.25]{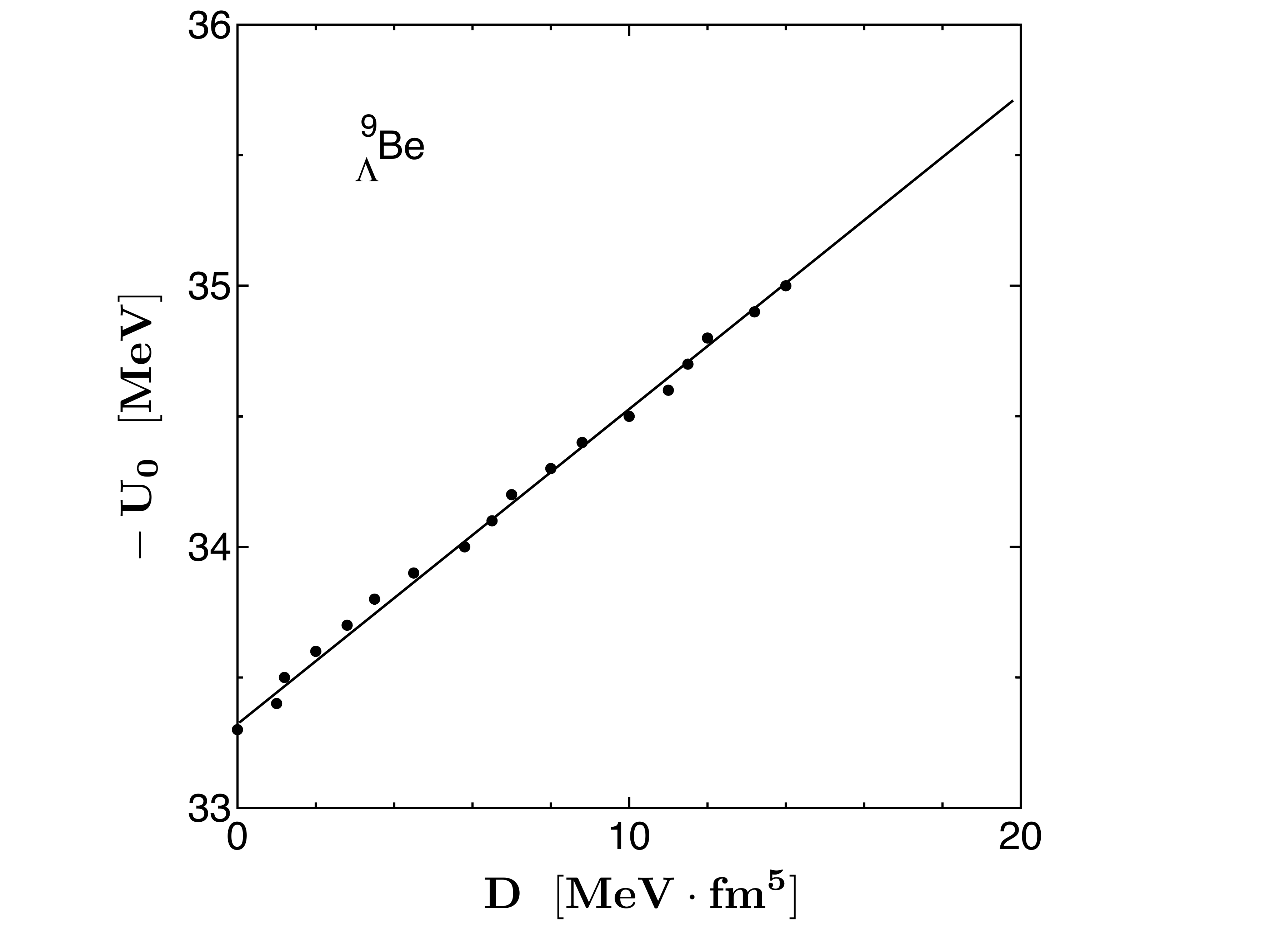}
\caption{Correlation between the $\Lambda$-nuclear potential depth $U_0$ and the 
strength $D$ of the surface term for $^{9}_\Lambda$Be. The straight line connects points that reproduce 
the empirical $\Lambda$ binding energy, $B_\Lambda = 6.71$ MeV, using the THSR cluster model wave function as described in the text. }
\label{fig:dfac-be9}
\end{center}
\end{figure}

Next, consider $^9_\Lambda$Be. The linear relationship between $U_0$ and $D$ is also observed 
in this case as shown in Fig. \ref{fig:dfac-be9}, but with a displacement towards smaller values of $D$ as compared to $^{13}_\Lambda$C. The nuclear core density of $^9_\Lambda$Be with its compressed distribution has a radius close to that of the core in $^{13}_\Lambda$C. 
Now a central potential depth $U_0 = - 33.3$\,MeV alone would give $B_\Lambda$ = 
6.24\,MeV, and the empirical $B_\Lambda$ = 6.71\,MeV is reached already by just adding the 
effective mass term. There appears to be no need for including an extra surface 
term unless a stronger central potential is preferred. Choosing $U_0 = - 
35$\,MeV, for example, would lead to overbinding, $B_\Lambda$ = 7.5\,MeV (including the effective mass 
correction of about 0.5 MeV). The empirical $B_\Lambda$ of $^9_\Lambda$Be is then 
reproduced with a surface coupling strength $D = 14$\,MeV$\cdot$fm$^5$. This might appear
to fall out of the range of surface coupling strengths, $D \simeq 40-50$ MeV$\cdot$fm$^5$, 
discussed previously for $^{13}_\Lambda$C and heavier hypernuclei.
A reason for this different behavior can be traced to the strong deformation of the $^9_{\Lambda}$Be 
ground state consisting of $2\alpha$ clusters and a $\Lambda$. A more detailed analysis 
would require taking this deformation explicitly into account in the calculation of the density 
profile, rather than using an angular average as in Eq. (\ref{eq:dsty}).

\begin{figure}[htbp]
\begin{center}
\includegraphics[scale=0.25]{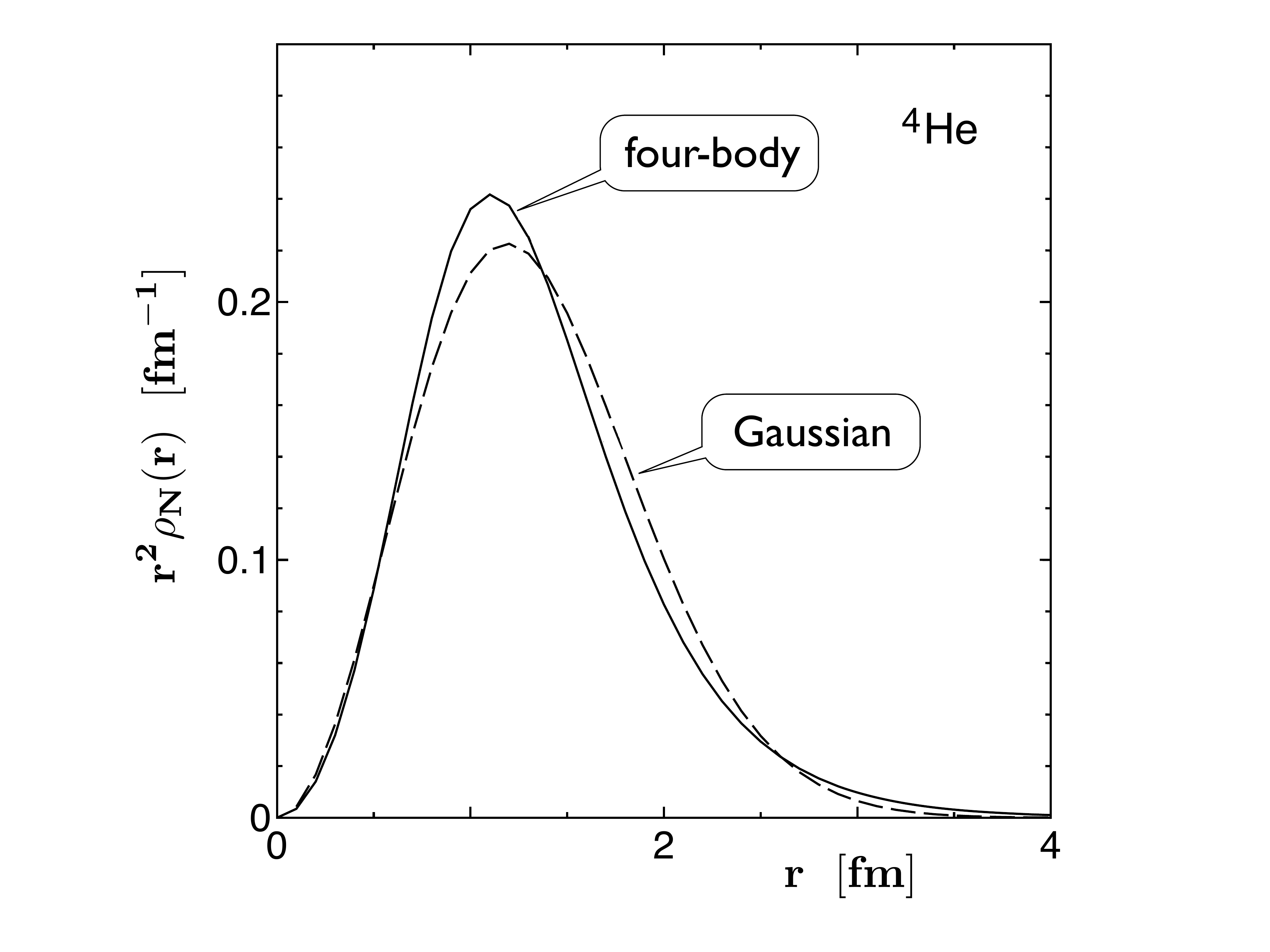}
\caption{Density profile (multiplied by $r^2$) of $^4$He derived from the realistic four-body calculation of Ref. \cite{HGK2004} (solid curve). This input
is used as the alpha core density in the calculation of the $^5_\Lambda$He binding energy. Also shown for
comparison is a standard Gaussian density profile (dashed curve).}
\label{fig:densityHe4}
\end{center}
\end{figure}

\begin{figure}[htbp]
\begin{center}
\includegraphics[scale=0.25]{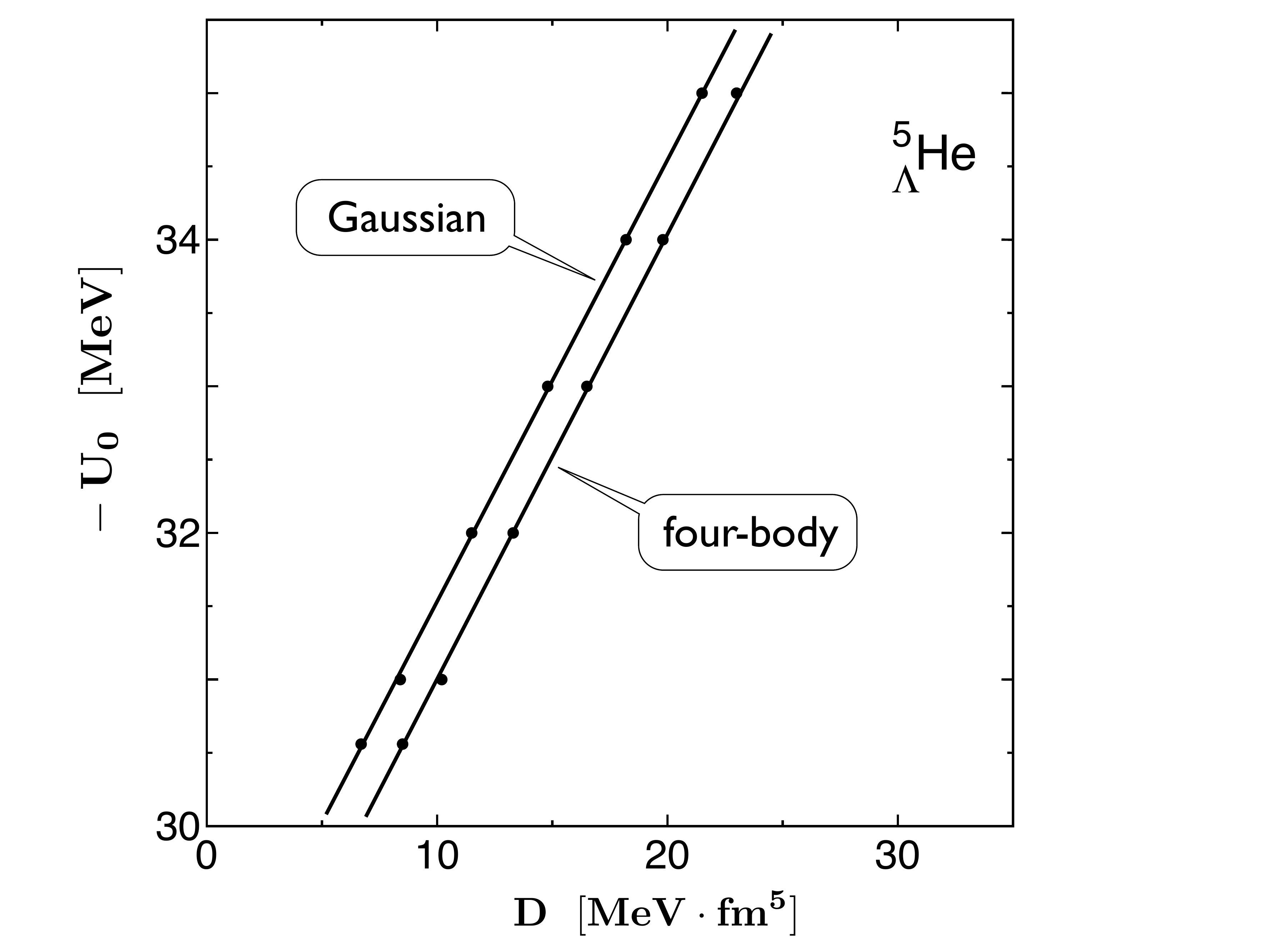}
\caption{Correlation between the $\Lambda$-nuclear potential depth $U_0$ and the 
strength $D$ of the surface term for $^{5}_\Lambda$He. The $^4$He core density (see Fig.\ref{fig:densityHe4}) is either of Gaussian form or taken from a microscopic four-body calculation \cite{HGK2004}. The straight lines connect points that reproduce the empirical $\Lambda$ binding energy, $B_\Lambda = 3.12$ MeV. }
\label{fig:dfac-he5}
\end{center}
\end{figure}

As a final point a brief discussion of the $^5_\Lambda$He hypernucleus is also instructive. This is a prototype 
system featuring the interaction of the $\Lambda$ hyperon with the $\alpha$ particle core. In principle, an 
{\it ab-initio} calculation of the $^5_\Lambda$He binding energy requires solving a five-body problem. 
Given the chiral SU(3) effective interaction between the $\Lambda$ and the core, this reduces to a two-body 
problem with the $^4$He core density distribution as input, assuming that this compact distribution does not change much in the presence of the hyperon. Here we use the density resulting from 
the microscopic four-body calculation \cite{HGK2004} that reproduces the $^4$He matter radius,
$\langle r^2\rangle_{m}^{1/2} = [\langle r^2\rangle_{ch} - \langle r^2\rangle_{p}]^{1/2} =1.45$ fm, derived from the empirical $^4$He charge radius $\langle r^2\rangle_{ch}^{1/2} = 1.68$ fm together with the proton charge radius $\langle r^2\rangle_{p}^{1/2} = 0.85$ fm. As mentioned, this density
profile differs from a simple Gaussian form that reproduces the same radius (see Fig.\ref{fig:densityHe4}). Calculating the $\Lambda$ binding energy 
and choosing again the potential parameters, $U_0$ and $D$, such as to reproduce the 
empirical $B_\Lambda$ = 3.12\,MeV for $^5_\Lambda$He, one finds once more a linear relationship
between $U_0$ and $D$ (see Fig.\ref{fig:dfac-he5}). Using $U_0 = - 35$\,MeV, 
a surface coupling  $D \simeq 23$\,MeV$\cdot$fm$^5$ would be suggested. The stronger correlation between $U_0$ and $D$ in the $^5_\Lambda$He case reflects the more compact $\alpha$ particle core in this light hypernucleus with its more pronounced surface gradient. The sensitivity with respect to the 
$\nabla^2\rho_N$ term becomes apparent by comparison with a standard Gaussian density. Of course,
for such light and compact hypernuclei, one reaches the limit of applicability for the gradient expansion Eq. (\ref{eq:lambdapot}) with a local density $\rho_N(r)$, and higher powers of gradients are expected to
become non-negligible. 

\section{Summary}

The present analysis of light hypernuclei using microscopic 
cluster model wave functions points to a sensitive interplay between the Hartree-type central 
$\Lambda$-nuclear potential and terms involving derivatives of the nuclear
core density, $\rho_N$, in the hypernucleus. These derivative terms have their 
well-founded origin in the in-medium $\Lambda$ self-energy derived from chiral SU(3) 
meson-baryon effective field theory. A part proportional to $\vec\nabla \cdot R(\rho_N)
\vec\nabla$ effectively increases the mass and reduces the kinetic energy of the 
$\Lambda$ in the hypernucleus, thereby increasing the $\Lambda$ binding energy. 
A repulsive surface term proportional to $\vec \nabla^2\rho_N$ counteracts this 
tendency. A systematic linear correlation is found between the strengths of the
central attraction and the surface repulsion. The results for $^{13}_\Lambda$C turn out 
to be consistent with earlier mean-field calculations using similar input. For the lightest
hypernuclei studied in this work ($^9_\Lambda$Be, $^5_\Lambda$He) the linear correlation 
just mentioned is also found but with a weaker surface term. For $^9_\Lambda$Be 
the loosely bound $2\alpha$ structure of the $^8$Be 
core nucleus, although compressed by the presence of the $\Lambda$ hyperon, makes 
this system special. The $^5_\Lambda$He case is presumably at the borderline of applicability
of the present approach, as explained.  

In summary, a significant result of the present study is
that an independent calculation of $^{13}_\Lambda$C using a microscopic wave function 
confirms the importance of the $\Lambda$-nuclear derivative coupling terms predicted
by in-medium chiral SU(3) effective field theory.

\section*{Appendix}
In this appendix we provide the analytical expressions for the strength functions 
$R(\rho_N)$ and $D(\rho_N)$ as obtained by evaluating the momentum-dependent 
in-medium $\Lambda$ self-energy in SU(3) chiral effective field theory up to 
two-loop order. 

The one-kaon exchange Fock diagram gives:
\begin{equation} M_\Lambda^{-1}R(\rho_N)^{(K)} = {({\cal D}+3{\cal F})^2 \over( 6 \pi 
f_\pi)^2 } {2m_K^2 k_F^3 \over (m_K^2+k_F^2)^2}\,, \end{equation} 
with the SU(3) axial vector couplings ${\cal D} =0.84$, ${\cal F} =0.46$ and the pion decay 
constant $f_\pi = 92.4\,$MeV. The Fermi momentum $k_F$ is related to the 
nuclear density by  $\rho_N = 2k_F^3/3\pi^2$.

The iterated pion-exchange diagram with a $\Sigma$ hyperon in the intermediate 
state gives:
\begin{eqnarray}  M_\Lambda^{-1}R(\rho_N)^{(2\pi)} &=& {{\cal D}^2 g_A^2 M_B m_\pi^2 \over 
24\pi^3f_\pi^4}\Bigg\{-4 \arctan{\sqrt{u}\over 2+\sqrt{4\delta-u}}
+ 
 {\sqrt{u} \over [u+(\delta-1)^2]^2} \bigg[2(1-\delta)^3 \nonumber \\
&&+u(3-\delta)  -{2\delta^3(1+\delta)^2\over \sqrt{4\delta-u}} +
\sqrt{4\delta-u}\,\bigg({1\over 2}(u^2-3u+\delta^4-5\delta^2) \nonumber \\
&& +u \delta^2+
2\delta^3+3\delta-1\bigg)\bigg]\Bigg\}\,,
\end{eqnarray} 
with the dimensionless variables $u =(k_F/m_\pi)^2$ and $\delta = (\Delta/m_\pi)^2$, 
where the small scale $\Delta = 285\,$MeV is related to the $\Sigma\Lambda$ mass 
splitting by $M_\Sigma-M_\Lambda = \Delta^2/M_B$. Furthermore, $g_A = {\cal D} + {\cal F}=1.3$ is the 
nucleon axial vector coupling constant and $M_B = 1047\,$MeV denotes an average 
baryon mass.

The same (long-range) two-pion exchange mechanism gives for surface coupling strength:
\begin{eqnarray} 
\rho_N D(\rho_N)^{(2\pi)} &=& {{\cal D}^2 g_A^2 M_B m_\pi^2 \over 
96\pi^3f_\pi^4}\Bigg\{4(3-4\delta+2u) \arctan{\sqrt{u}\over 2+\sqrt{4\delta-u}}
+ {\sqrt{u} \over [u+(\delta-1)^2]^2} \nonumber \\ 
&& \hspace{-2.5cm} \times \bigg[-{15 u^2 \over 2}+
{u\over 6}(\delta^3-90\delta^2+180 \delta -77) +2(3-4\delta) (\delta-1)^3-{\delta^3
(1+\delta)^2\over \sqrt{4\delta-u}} \nonumber \\ 
&& \hspace{-3.0cm} +\sqrt{4\delta-u}\,
\bigg({25u^2 \over 6}+{u\over 12} (99\delta^2-188 \delta+78)
+ {\delta^2\over 4}(17\delta^2-58\delta+85)-13\delta+3\bigg)\bigg]\Bigg\}
\,.
\end{eqnarray}
In addition there are (small) Pauli-blocking corrections which reduce the 
contributions $R(\rho_N)^{(2\pi)}$ and  $D(\rho_N)^{(2\pi)}$ with increasing density.  

\section*{Acknowledgements}
This work has been partially supported by a cooperation agreement between the RIKEN Nishina Center and ECT*,
by BMBF, and by DFG through CRC110 ''Symmetries and the Emergence of Structures in QCD". E.H. thanks the 
Yamada Science Foundation for support. One of the authors (W.W.) is grateful to Tetsuo Hatsuda and Emiko Hiyama for their kind hospitality at RIKEN. The numerical calculations were performed on
the HITACHI SR16000 at YITP and KEK.



\begin{thebibliography}{99}

\bibitem{Tamura2010}
H. Tamura, Prog. Theor. Phys. Suppl. {\bf 185}, 315 (2010); 
O. Hashimoto and H. Tamura, Prog. Part. Nucl. Phys. {\bf 57}, 564 (2006), and 
references therein.

\bibitem{Hiyama2009}
E. Hiyama and T. Yamada, Prog. Part. Nucl. Phys. {\bf 63}, 339 (2009).

\bibitem{Gal2008}
A. Gal and R.S. Hayano (Eds.), {\it Special Issue on Strangeness Nuclear Physics}, Nucl. Phys. {\bf A804} (2008). 

\bibitem{Millener2008}
D.J. Millener, Nucl. Phys. {\bf A804}, 84 (2008); Nucl. Phys, {\bf A881}, 298 (2012).

\bibitem{Tamura2012}
H. Tamura, M. Ukai, T.O. Yamamoto, and T. Koike, Nucl. Phys, {\bf A881}, 310 (2012).

\bibitem{KW2005}
N. Kaiser and W. Weise, Phys. Rev. {\bf C71}, 015203 (2005).

\bibitem{FKVW2009}
P. Finelli, N. Kaiser, D. Vretenar, and W. Weise, Phys. Lett. {\bf B658}, 90 (2007), Nucl. Phys. {\bf A831}, 163 (2009).

\bibitem{WT1977}
K. Wildermuth, and Y. C. Tang, {\it A Unified Theory of the Nucleus} (Vieweg, Braunschweig, Germany, 1977).

\bibitem{IHS1980}
K. Ikeda, H. Horiuchi, and S. Saito, Prog. Theor. Phys. Suppl. {\bf 68}, 1 (1980).

\bibitem{epelbaum2012}
E. Epelbaum, H. Krebs, T. L\"uhde, D. Lee, and U.-G. Mei{\ss}ner, Phys. Rev. Lett. {\bf 109}, 252501 (2012).

\bibitem{thsr}
A. Tohsaki, H. Horiuchi. P. Schuck and G. R\"{o}pke, Phys. Rev. Lett. {\bf 87}, 192501 (2001).

\bibitem{funaki_12C}
Y. Funaki, A. Tohsaki, H. Horiuchi, P. Schuck, and G. R\"{o}pke, Phys. Rev. C {\bf 67}, 051306(R) (2003).

\bibitem{funaki_concept}
Y. Funaki, H Horiuchi, W. von Oertzen, G. R\"{o}pke, P. Schuck, A. Tohsaki, and T. Yamada, 
Phys. Rev. C {\bf 80}, 064326 (2009).

\bibitem{YFHIT2008} 
T. Yamada, Y. Funaki, H. Horiuchi, K. Ikeda, and A. Tohsaki, Prog. Theor. Phys. {\bf 120}, 1139 (2008).

\bibitem{Win2008}
 M.T. Win and K. Hagino,  Phys. Rev. C {\bf 78}, 054311 (2008).

\bibitem{Isaka2011}
M. Isaka, M. Kimura, A. Dote, and A. Ohnishi, Phys. Rev. C {\bf 83}, 044323 (2011).

\bibitem{HGK2004}
 E. Hiyama, B.F. Gibson, and M. Kamimura, Phys. Rev. C {\bf 70}, 031001(R) (2004).

\bibitem{volkov}
A. B. Volkov, Nucl. Phys. {\bf{74}}, 33 (1965).

\bibitem{kamimura}
 Y. Fukushima and M. Kamimura, {\it{Proc. Int. Conf. on Nuclear Structure}}, 
 Tokyo, 1977, ed. T. Marumori (Suppl. of J. Phys. Soc. Japan, Vol.44, 1978), 
 p.225; M. Kamimura, Nucl. Phys. {\bf A 351}, 456 (1981).
 
 \bibitem{HKM1997}
E. Hiyama, M. Kamimura, T. Motoba, T. Yamada, and Y. Yamamoto, Prog. Theor. Phys. {\bf 97}, 881 (1997).

\bibitem{expBe}
M. Juric {\it et al.}, Nucl. Phys. {\bf B52}, 1 (1973).

\bibitem{expC}
M. Canywell {\it et al.}, Nucl. Phys. {\bf A236}, 445 (1974).

\bibitem{Haidenbauer2013}
J. Haidenbauer, S. Petschauer, N. Kaiser, U.-G. Mei{\ss}ner, A. Nogga, and W. Weise, Nucl. Phys. {\bf A915}, 24 (2013).

\bibitem{funaki_8Be}
Y. Funaki, H. Horiuchi, A. Tohsaki, P. Schuck, and G. R\"{o}pke, Prog. Theor. Phys. {\bf 108}, 297 (2002).

\bibitem{zhou_bo}
B. Zhou, Z. Z. Ren, C. Xu, Y. Funaki, T. Yamada, A. Tohsaki, H. Horiuchi, P. Schuck, and G. R\"{o}pke, Phys. Rev. C {\bf 86}, 014301 (2012).

\bibitem{carbon}
Y. Fujiwara, Y. Suzuki, H. Horiuchi, K. Ikeda, M. Kamimura, K. Kat\={o}, Y. Suzuki, and E. Uegaki, Prog. Theor. Phys. Suppl. {\bf 68}, 29 (1980).

\bibitem{hiyama2000}
E. Hiyama, M. Kamimura, T. Motoba, T. Yamada, and Y. Yamamoto, Phys. Rev. Lett. {\bf 85}, 270 (2000).

\bibitem{GHW}
J. J. Griffin and J. A. Wheeler, Phys. Rev. {\bf 108}, 311 (1957).

\bibitem{devries1987}
H. De Vries, C.W. De Jager, and C. De Vries, in: {\it Atomic Data and Nuclear Data Tables} {\bf 01} (1987);
DOI:10.1016/0092-640X(87)90013-1.

\bibitem{Yamada85}
T. Yamada, T. Motoba, K. Ikeda, and H. Band\={o}, Prog. Theor. Phys. Suppl. {\bf 81}, 104 (1985).
\end{thebibliography}
\end{document}